\documentclass[prb,showpacs,showkeys,superscriptaddress,footnote]{revtex4}
\usepackage{amsfonts}
\usepackage{amsmath}
\usepackage{amssymb}
\usepackage{graphicx}

\begin{document}

\title{Covariant path integral formalism of relativistic quantum mechanics
along proper time}
\author{H. Y. Geng}
\affiliation{Department of Quantum Engineering and Systems Science, The University of Tokyo, Hongo 7-3-1, Tokyo 113-8656,
Japan}

\begin{abstract}
A space-time symmetric and explicitly Lorentz covariant path integral
formalism of relativistic quantum mechanics is proposed, which produces partial locally
correlations of quantum processes of massive particles with the velocity of light
at low energy limits. A superluminal correlation is also possible if
anti-particles that moving along reverse time direction are excited.
This provides
a new point of view to interpret EPR experiments, also leaks a light of
hope for hidden variable theories.
\end{abstract}

\volumeyear{year}
\volumenumber{number}
\issuenumber{number}
\eid{identifier}
\maketitle


\section{Introduction}
\label{sec-Intro}

In spite of quantum mechanics has demonstrated the amazing power to describe micro-phenomena,
its behavior is so different from macro-intuitions that attempts to recover (quasi-) deterministic
movement of particles never give up.\cite{bohm52,nelson66,olavo99} However, due to
non-local correlation of quantum states, these attempts all suffer quite difficulties to achieve
self-consistent results and received a despaired strike by Aspect's experiments\cite{aspect81,aspect82}
on EPR correlations.\cite{epr35} Many evidences have revealed that non-locality (or entanglement) is a fundamental
property of quantum process, and to restore the motion of particles back to deterministic
manner is in fact impossible.

Nevertheless, one may ask a further question: is such kind of non-local correlation globally (i.e., simultaneously
around the whole space) correlated or partially localized? From a point of view of standard
non-relativistic quantum mechanics, since wave function is defined on the whole space, as well as all operators,
the theory must be globally correlated. In terms of Feynman's path integral
formalism, that is to say that when calculating the transition function between two quantum states
\begin{eqnarray}
\langle q_{t_{b}},t_{b}|q_{t_{a}},t_{a}\rangle &=& \int_{q\left( t_{a}\right)=q_{t_{a}}}^{q\left( t_{b}\right)=q_{t_{b}}}\mathcal{D}[q(t)]\mathcal{D}[p(t)] \nonumber\\
&&\times \exp\left[\frac{i}{\hbar}\int_{t_a}^{t_b}\mathrm{d}t\left\{\sum_{r} p_{r}(t)\frac{\mathrm{d}}{\mathrm{d}t}q_{r}(t)-H\left(q(t),p(t)\right)\right\}\right],
\label{eq:fpi}
\end{eqnarray}
all paths across the space are allowed,\cite{feynman48,masujima00} which results in globally and instantaneous corrections. Here path integral $\int_{q\left(
t_{a}\right)=q_{t_{a}}}^{q\left(
t_{b}\right)=q_{t_{b}}}\mathcal{D}[q(t)]\mathcal{D}[p(t)]$ means to sum
over the whole coordinate and momentum spaces along time-sliced paths with fixed end
points at $q_{t_{a}}$ and $q_{t_{b}}$ (see chapter 1 of Ref.~\onlinecite{masujima00} for details).
Thus one may wonder how can we separate an atomic (quantum) system from the universe
and treat it isolatedly as we did heaps of times before? A
consistent theory should explain why to ignore effects due to
supernova explosions or other processes occurring at somewhere in the cosmos is
reasonable. It is difficult within the framework of standard quantum mechanics
\footnote{In wave dynamics, since wave functions are defined on the
whole space, when considering an atomic system, say, hydrogen atom, then the
Schr{\"o}dinger equation should also include terms corresponding to other ions distributing
across the universe (even charged black holes and strong magnetic fields).
many of these ions have much larger charges then proton, as well as larger
contributions to system energy. In this situation, the effects due to
proton can be neglected, and one cannot obtain the well-known quantum
properties of hydrogen atom, which has been confirmed by experiments many times.
To aviod this difficulty, one has to assume that electron is
restricted locally within some spatial region and satisfies certain isolation assumption. But it
is imcompatible with the postulations of standard
quantum mechanics, as well as the boundary conditions.}
Within the framework of path integral formalism, one may would like to employ the
stationary phase condition to argue that contributions from those paths
with large deviation from classical one's canceled completely and results in required isolation conditions. But notice (a)
this is only true for quasi-classical approximations; (b)
obviously different paths would have different variations of the contribution due to
changes of the dynamic environment along each course, there is no guarantee that all of these
variations with time can be completely and simultaneously
canceled all the time. In a sense of that, we reach a point that some paths in Eq.(\ref{eq:fpi})
should be forbidden in a self-consistent quantum theory in order to satisfy the isolation hypothesis and
to remove instantaneous correlations.

Actually, it will be seen in this paper that to generalize the Feynman's
formalism slightly to covariant relativistic case will produce partial locality to quantum mechanics. This property is crucial to ensure that
experiments on atomic system are really corresponding to the behavior of that
system but not those relating to the external cosmos.
It is necessary to point out that several attempts to set up a relativistic particle path integral
have been proposed before,
but they failed to produce the required isolation condition.
The employed formalisms are rather unsatisfactory due to lack of explicit
Lorentz covariance and treated time and spatial coordinates on different footings.\cite{Felea01,Pazma79,kull99,Gosselin98,Padmanabhan94,Fanchi05}
The physical implications associated with the derivations also are not clear
enough.
Therefore these attempts are in fact a kind of mathematical techniques for
conventional quantum mechanics, and cannot be taken as the foundation for
developing a new theory.
In subsequent sections, a variational principle on world lines is developed,
which is necessary in order to generalize Feynman's formalism to Lorentz covariance. The relationship with standard quantum mechanics
is discussed in section \ref{sec-cpi}. Some understandings arised from this
new point of view are given, followed by a comparison with previous derivations.

\section{Principle of formal action}
\label{sec-p-action}

Thanks to the theory of relativity, time is deprived its privilege in
motion equations and became an ordinary dimension of space-time. If one still
would like to use Lagrangian variational principle to recover the position
of action principle in physics, finding out an other parameter to take the place of
time in Newtonian mechanics is necessary. Considering the movement of a particle with non-zero rest mass in space-time,
we always can choose a frame of reference in which the particle is at rest. Usually it
is non-inertial. However, it is reasonable to assume that instantaneously
one can employ an inertial frame to approximate it (in terms of the general theory of
relativity, the space-time manifold can be approximated by a series of Minkowski spacetimes locally\cite{weinberg72}). In this way the movement
of a particle can be described by a series of inertial frames, which
relating to the observer by respective Lorentz transformations\footnote{If all interactions are instantaneous (as well recognised),
non-inertial reference frame loses its intrinsic physical meaning and this quasi-inertial approximation becomes exact.}
\begin{equation}
x_{\mu}(\beta)=\sum_{\nu}\Lambda_{\mu}^{\phantom{\mu}\nu}(\beta)x^{0}_{\nu}.
\label{eq:Lorentz-t}
\end{equation}
Here $\beta$ is the ratio of instantaneous velocity to light speed, i.e., $v/c$.
Under these transformations, the length of proper time $\tau$ is invariant and satisfies
$c^{2}(\mathrm{d}\tau)^{2}=\sum_{\mu,\nu}\eta^{\mu\nu}\mathrm{d}x_{\mu}\mathrm{d}x_{\nu}$,
where $\eta^{\mu\nu}$ is the metric of Minkowski space.
Usually $\tau$ can be expressed as a functional as $\tau=\tau(\beta,x_{\mu})$.
Thus the variation of frames characterized by parameter $\beta$ can also
be characterized by proper time $\tau$. Namely, with Eq.(\ref{eq:Lorentz-t})
we have $x_{\mu}=x_{\mu}(\tau)$. In this way the proper time takes the
privilege position of time as in Newtonian mechanics, and the motion equations
of $x_{\mu}$ along $\tau$ can be obtained by the generalized principle of
least action that defined on world lines.\footnote{For massless particles, Eq.(\ref{eq:Lorentz-t}) fails
and the equations of motion is independent from proper time. It corresponds
to a static case in Newtonian framework.}

Suppose a relativistic dynamical system that can be determined by a characteristic functional $L(x_{\mu},\dot{x}_{\mu})$,
where $\dot{x}_{\mu}=\mathrm{d}x_{\mu}/\mathrm{d}\tau$, then one may
have a formal action defined on world lines
\begin{equation}
S\left[x_{\mu}(\tau)\right]=\int_{\tau_{i}}^{\tau_{f}}L(x_{\mu},\dot{x}_{\mu})\mathrm{d}\tau
\label{eq:action-1}
\end{equation}
with fixed end points at $x_{\mu|\tau_{i}}$ and $x_{\mu|\tau_{f}}$.
The variation of this action with respect to world lines with fixed boundary
conditions $\delta x_{\mu}(\tau_{i})=\delta x_{\mu|\tau_{i}}=0$ and $\delta x_{\mu}(\tau_{f})=\delta x_{\mu|\tau_{f}}=0$
gives
\begin{equation}
\frac{\mathrm{d}}{\mathrm{d}\tau}\left(\frac{\partial L}{\partial \dot{x}_{\mu}}\right)-\frac{\partial L}{\partial x_{\mu}}=0
\label{eq:c-LE}
\end{equation}
when $\delta S[x_{\mu}(\tau)]=0$. This is just the generalized covariant
formalism of least action principle, and one can employ Eq.(\ref{eq:c-LE})
to reproduce the covariant motion equation of particles.

Using the formal conjugate momentum $p^{\mu}=\partial L/\partial \dot{x}_{\mu}$,
we can define a formal Hamiltonian by a Legendre transform
\begin{equation}
M\left(x_{\mu},p^{\mu}\right)=\sum_{\mu}p^{\mu}\dot{x}_{\mu}-L(x_{\mu},\dot{x}_{\mu}).
\label{eq:M}
\end{equation}
Usually it is Lorentz invariant. Then the equation of motion can also be given by
\begin{equation}
\dot{x}_{\mu}=\frac{\partial M}{\partial p^{\mu}},\ \ \dot{p}^{\mu}=-\frac{\partial M}{\partial x_{\mu}}.
\label{eq:M-meq}
\end{equation}
It is evident from above discussion that if $M$ is not explicitly
$\tau$-dependent, then the formal Hamiltonian is a conservative quantity.
When $x_{\mu}$ and $p^{\mu}$ varied independently, the formal action of Eq.(\ref{eq:action-1})
can be expressed as
\begin{equation}
S\left[x_{\mu}(\tau),p^{\mu}(\tau)\right]=\int_{\tau_{i}}^{\tau_{f}}\left[\sum_{\mu}p^{\mu}\dot{x}_{\mu}-M\left(x_{\mu},p^{\mu}\right)\right]\mathrm{d}\tau,
\label{eq:action-2}
\end{equation}
which is Lorentz invariant too.

For a free particle, one has $\mathrm{d}p^{\mu}/\mathrm{d}\tau=0$ for
4-momentum $p^{\mu}$.\cite{weinberg72} It is easy to prove that the corresponding formal Hamiltonian
should be
\begin{equation}
M=\frac{c}{2}\sqrt{p^{\mu}p_{\mu}}.
\label{eq:form-H}
\end{equation}
Then one gets $M=m_{0}c^{2}/2$ where $m_{0}$ is the rest mass because $\sum_{\mu}p^{\mu}p_{\mu}=m_{0}^{2}c^{2}$.
The conservation law of $M$ in fact becomes the conservation of rest mass or
rest energy. In this formalism we see that mass no longer appears as an \emph{aprior} parameter but
a dynamical variable analogous to Hamiltonian in classical physics.

\section{Covariant path integral}
\label{sec-cpi}
As discussed in introduction section, self-consistent quantum mechanics should be at
least partial localized. If it is true, there is no any reason to save the conception
that wave function (or quantum state) exclusively describes just the motion of that particles.\footnote{Partial localization inevitably leads to a concept of correlation
lenghth between quantum processes. Owing to the law of causality, measuments beyond correlation
length on the same state should be independent, which will result in logical contradiction
if one insists, say, single particle wave function just describes the motion
of only \emph{that} particle, beacuse independent measurments will give
an odd conclusion such that a particle may appear at different places in spatial space simultaneously.}
This clew of thought eventually leads to a theory of quantum fields. However,
here we would like to restrain our discussions within quantum mechanics.

\begin{figure}[th]
\centerline{\includegraphics*[width=2.4in]{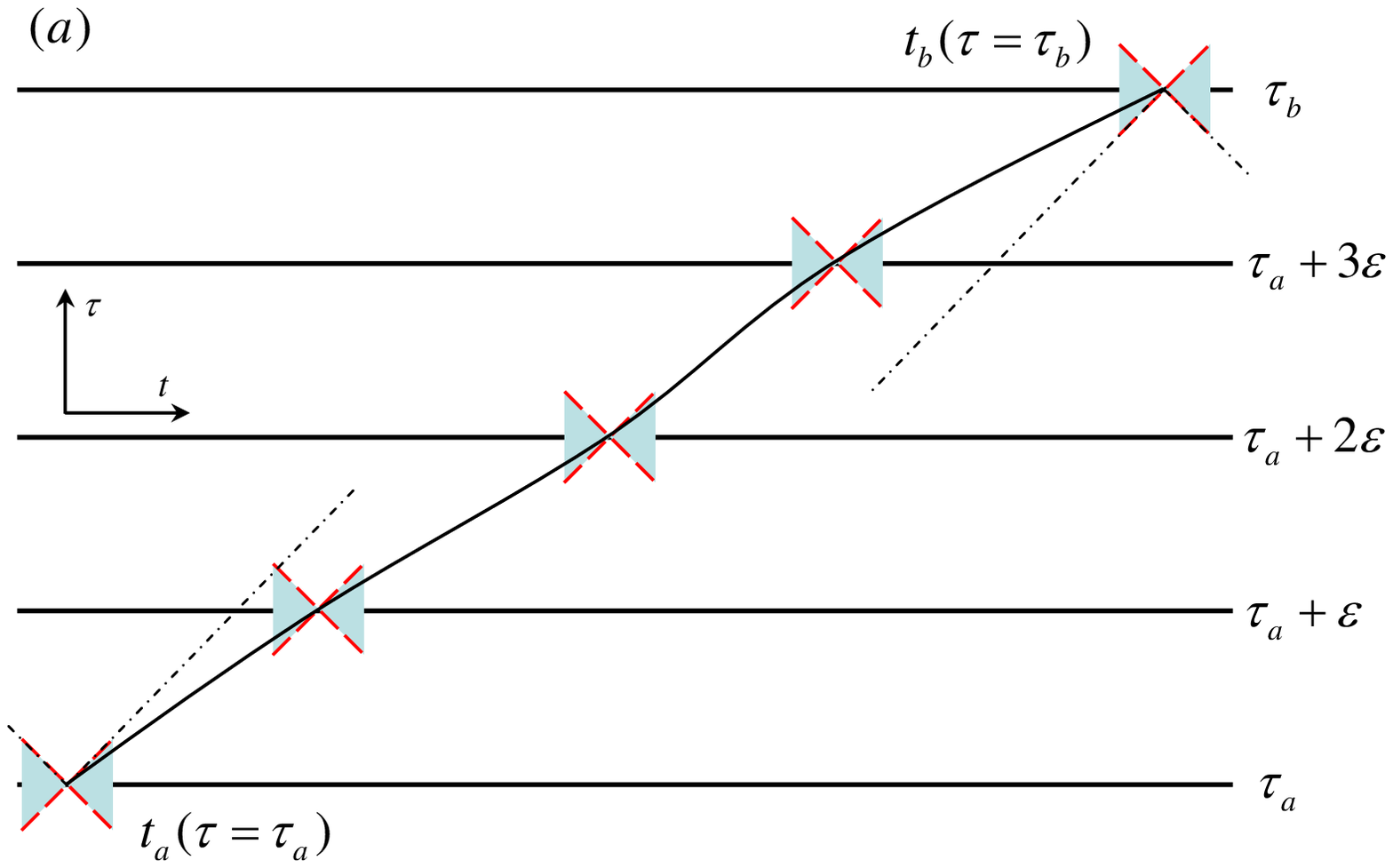}
\includegraphics*[width=1.8in]{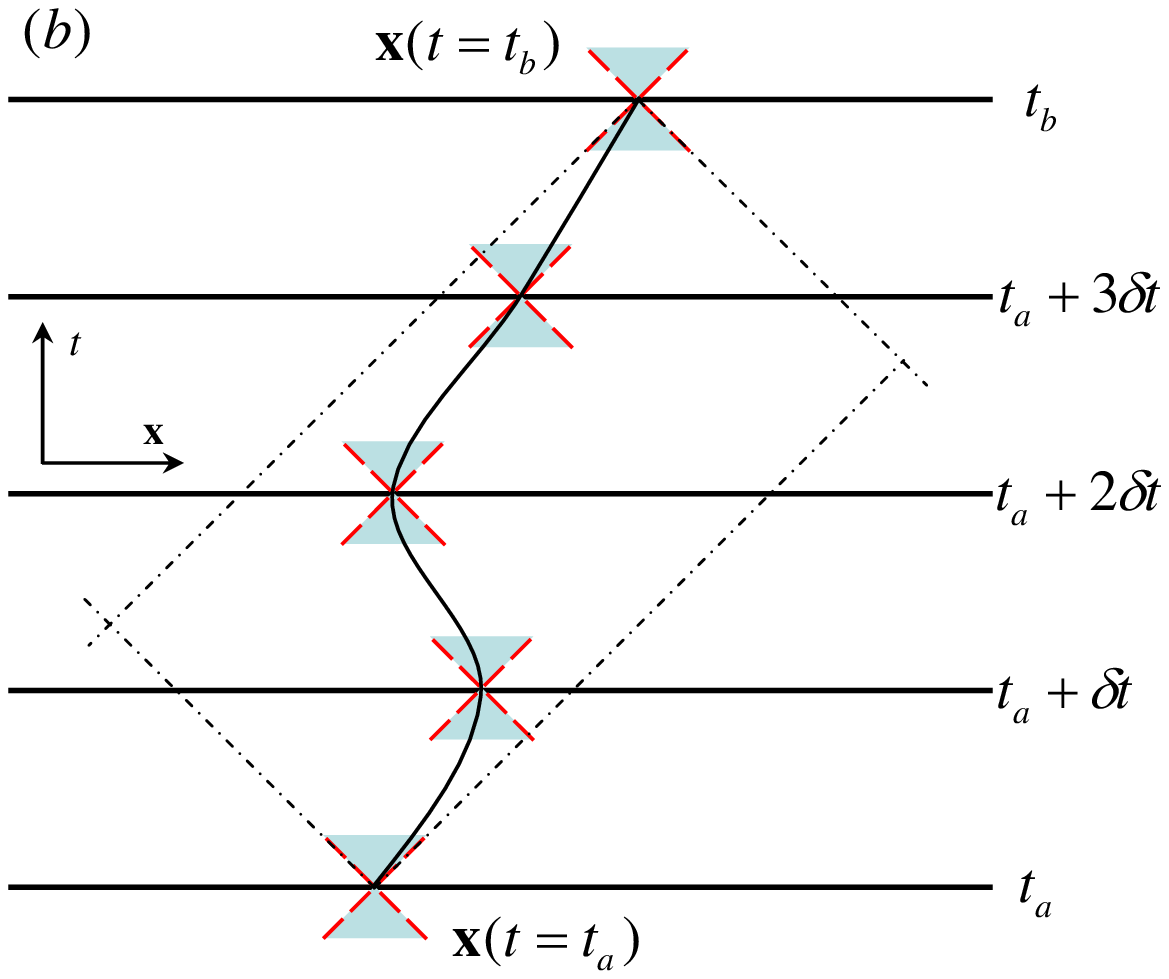}}
\vspace*{8pt}
\caption{Projections of allowed paths in covariant path integral formalism in (1+1) space-time.
  (a) proper time slices for Eq.(\ref{eq:c-fpi}) where paths must satisfy $|\mathrm{d}\tau/\mathrm{d}t|\leq 1$ (domain $\mathbb{C}_{1}$);
  (b) spatial summation lattice between time {$t_{a}$} and {$t_{b}$} (domain $\mathbb{C}_{2}$),
  not only all paths belonging to end points {$\mathbf{x}(t_{a})$} and {$\mathbf{x}(t_{b})$}
  are restricted by the dot-dashed rectangle, every path also should be within the
  instantaneous light cones all the time to ensure a properly definition of proper time.
  \protect\label{fig:cpi}}
\end{figure}

Following Feynman's consideration, it is quite natural to assume that
the transition probability amplitude between two quantum states is
completely determined by a phase factor given by the formal action of particle Eq.(\ref{eq:action-2}), which results in the covariant
version of Eq.(\ref{eq:fpi})
\begin{align}
\langle x_{\mu}(\tau_{b}),&\tau_{b}|x_{\mu}(\tau_{a}),\tau_{a}\rangle=\int_{x_{\mu}\left( \tau_{a}\right)=x_{\mu|\tau_{a}}}^{x_{\mu}\left( \tau_{b}\right)=x_{\mu|\tau_{b}}}
\mathcal{D}[x(\tau)]\mathcal{D}[p(\tau)] \nonumber\\
&\times \exp\left[\frac{i}{\hbar}\int_{\tau_a}^{\tau_b}\mathrm{d}\tau\left\{\sum_{\mu,\nu}p^{\nu}(\tau)\frac{\mathrm{d}}{\mathrm{d}\tau}x_{\mu}(\tau)\delta_{\mu\nu}-M\left(x_{\mu}(\tau),p^{\mu}(\tau)\right)\right\}\right].
\label{eq:c-fpi}
\end{align}
Evidently, this is just a natural special relativity generalization of Feynman's path integral formalism.
The proper time sliced expression is given by (differs from other path integral formalisms
where a time-slicing scheme is used)
\begin{align}
\langle x_{\mu}(\tau_{b}),&\tau_{b}|x_{\mu}(\tau_{a}),\tau_{a}\rangle=\lim_{n\rightarrow\infty \atop n\cdot\delta\tau=\tau_{b}-\tau_{a}}
\int_{\mathbb{C}}\prod_{k=1}^{n-1}\frac{\mathrm{d}^{4}x_{\tau_{k}}}{\left(2\pi\hbar\right)^{2}}
\int\prod_{k=0}^{n-1}\frac{\mathrm{d}^{4}p_{\tau_{k}}}{\left(2\pi\hbar\right)^{2}}\nonumber\\
&\times \exp\left[\frac{i}{\hbar}\sum_{k=0}^{n-1}\left\{\sum_{\mu,\nu}p^{\nu}(\tau_{k})\delta x_{\mu}(\tau_{k})\delta_{\mu\nu}-\delta\tau\cdot M\left(x_{\mu}(\tau_{k}),p^{\mu}(\tau_{k})\right)\right\}\right],
\end{align}
where $\delta x_{\mu}(\tau_{k})=x_{\mu}(\tau_{k+1})-x_{\mu}(\tau_{k})$ and the
integral domain $\mathbb{C}$ of space-time coordinates is not only determined by
boundary conditions $x_{\mu}(\tau_{b})=x_{\mu|\tau_{b}}$ and $x_{\mu}(\tau_{a})=x_{\mu|\tau_{a}}$,
it also has to satisfy a condition of that $\delta\tau$ should be defined properly which constrained
by physical conditions $|\mathrm{d}\tau/\mathrm{d}t|\leq 1$ and $\sum_{\mu,\nu}\eta^{\mu\nu}\mathrm{d}x_{\mu}\mathrm{d}x_{\nu}\geq 0$.
Thus we have $\mathbb{C}=\mathbb{C}_{1}\bigcup\mathbb{C}_{2}$,
corresponding to an intrinsic property of Minkowski space-time and the non-spatial like motions of particle, respectively.
It is clear that the most significant difference between this and the
conventional path integral formalism lies on that in the latter case, all paths across the space are
possible, while Eq.(\ref{eq:c-fpi}) allows just some special paths,
as showed in Fig.\ref{fig:cpi}. For simplicity, only (1+1)-dimension case is shown here,
where time-component of proper time sliced paths are restricted by $|\mathrm{d}\tau/\mathrm{d}t|\leq 1$ and
spatial component belonging to end points {$\mathbf{x}(t_{a})$} and {$\mathbf{x}(t_{b})$}
are not only limited within the dot-dashed rectangle (relativistic causality), but also each path should
be time-like, i.e., contained in the instantaneous light cones all the time, to
guarantee $(\mathrm{d}\tau)^{2}\geq 0$ at classical limit (see Eq.(\ref{eq:c-fpi})). Evidently, this property recovers the global causality of relativity and partial locality to
the theory, the latter eventually leads to the isolation condition for
massive particles. It is easy to see that Eq.(\ref{eq:c-fpi}) contains conventional quantum
mechanics as approximation as Feynman case shown.
The superposition and composition law of the transition probability amplitude
is stated in a form of
\begin{equation}
  \int_{x_{\mu}\left( \tau_{a}\right)}^{x_{\mu}\left( \tau_{b}\right)}\mathcal{D}[x]\mathcal{D}[p]=\int_{\mathbb{C}}\mathrm{d}^{4}x'\int_{x'_{\mu}\left( \tau'\right)}^{x_{\mu}\left( \tau_{b}\right)}\mathcal{D}[x_{\uppercase \expandafter {\romannumeral 2} }]\mathcal{D}[p_{{\uppercase \expandafter {\romannumeral 2} }}]\int_{x_{\mu}\left( \tau_{a}\right)}^{x'_{\mu}\left( \tau'\right)}\mathcal{D}[x_{{\uppercase \expandafter {\romannumeral 1} }}]\mathcal{D}[p_{{\uppercase \expandafter {\romannumeral 1} }}].
\end{equation}
It is worthwhile to note that J. Schwinger once introduced proper time
depended Green functions to trick the gauge invariant problem,\cite{schwinger51} which
is different from our considerations here completely.

Consider the transition function of Eq.(\ref{eq:c-fpi}) with an infinitesimal increase in proper time $\tau$,
the wave function can be simplified as
\begin{align}
\psi(x_{\mu},\tau_{0}+\varepsilon)=&\int_{\mathbb{C}} \mathrm{d}^{4}x^{0}\int\frac{\mathrm{d}^{4}p}{(2\pi\hbar)^{4}}\,\psi(x^{0}_{\mu},\tau_{0})\nonumber\\
&\times\exp\left[\frac{i}{\hbar}\left\{\sum_{\mu,\nu}p^{\nu}
(x_{\mu}-x^{0}_{\mu})\delta_{\mu\nu}-\varepsilon\cdot M\left(x_{\mu},p^{\mu}\right)\right\}\right],
\label{eq:c-fpi-small-t}
\end{align}
with $\psi(x_{\mu})=\langle x_{\mu}|\psi\rangle$. As matrix element of operator $O(\hat{x}_{\mu}(\tau))$ is given by\cite{masujima00}
\begin{equation}
\langle x_{\mu|\tau_{b}}{\mid} O(\hat{x}_{\mu}(\tau)){\mid} x_{\mu|\tau_{a}}\rangle=
\int_{\mathbb{C}}
\mathcal{D}[x(\tau)]\mathcal{D}[p(\tau)]O(x_{\mu}(\tau))\exp\left[\frac{i}{\hbar}S(\tau_{b},\tau_{a})\right],
\end{equation}
it is quite easy to prove that the corresponding form of operators acting on wave function Eq.(\ref{eq:c-fpi-small-t}) are
\begin{equation}
i\hbar\frac{\partial}{\partial\tau}\Leftrightarrow \hat{M},\ \ -i\hbar\frac{\partial}{\partial x_{\mu}}\Leftrightarrow \hat{p}^{\mu}
\label{eq:xt-Operat}
\end{equation}
to the first order of $\varepsilon$ when $\varepsilon$ approaches zero.
Note here that the $4$-momentum operators have an oppositive sign comparing to traditional definition,
but it does not matter. Using Eq.(\ref{eq:xt-Operat}) we have the motion equation for wave function as
\begin{equation}
i\hbar\frac{\partial}{\partial\tau}\Psi=\hat{M}\Psi.
\end{equation}
For a formal Hamiltonian eigenstate $i\hbar\partial \Psi/\partial \tau=m\Psi$, one gets
\begin{equation}
\hat{M}\left(\partial/\partial x_{\mu}\right)\psi=m\psi,
\label{eq:eigeq-M}
\end{equation}
with $\Psi=\psi\exp(-im\tau/\hbar)$.

If studied system only includes gauge interactions, then the formal Hamiltonian of particle
part still keeps the form of Eq.(\ref{eq:form-H}) except for a slight modification $p^{\mu}\rightarrow p^{\mu}+A^{\mu}$.
So generally we can use the form of Eq.(\ref{eq:form-H}) for discussion. Applying
one more $\hat{M}$ on Eq.(\ref{eq:eigeq-M}) and making use of Eqs.(\ref{eq:form-H},\ref{eq:xt-Operat})
and $m=m_{0}c^{2}/2$, we reach the Klein-Gordon equation.
A detailed prove is given in appendix A to show this formal inference is true.
In order to obtain a linear equation, let's consider a mapping from $4$-vector to a Lorentz scalar
$x_{\mu}\rightarrow x=\sum_{\mu}\gamma^{\mu}x_{\mu}$. Conservation of the length
requires $\{\gamma^{\mu},\gamma^{\nu}\}=2\eta^{\mu\nu}$ with anticommutation
defined by $\{A,B\}=AB+BA$. Obviously $\gamma^{\mu}$ are exactly the Dirac matrices. In this
sense $\gamma^{\mu}/\sqrt{2}$ form a set of orthonormal basis with inner product given by anticommutation operation,
and $4$-vector $x_{\mu}$ can be viewed as the components of Lorentz scalar $x$ on this basis set $x_{\mu}=\{x,\gamma_{\mu}\}/2$.
Then the formal action can be reexpressed as
\begin{equation}
S\left[x(\tau),p(\tau)\right]=\int_{\tau_{i}}^{\tau_{f}}\left[\frac{1}{2}\{p,\dot{x}\}-M\left(x,p\right)\right]\mathrm{d}\tau,
\label{eq:action-3}
\end{equation}
with $M$ in Eq.(\ref{eq:form-H}) given by $M=pc/2=\sum_{\mu}\gamma_{\mu}p^{\mu}c/2$, which is
a Lorentz scalar but usually is a matrix too.
Then the Dirac equation is obtained directly from Eq.(\ref{eq:eigeq-M}).
A derivation of non-relativistic approximation is given in appendix B, which
reproduced the standard non-relativistic path internal formalism for free particles,
demonstrating the validity of our proposal.

By comparing with previous attempts to the same issue, our proposed scheme
is more systematic and elegant with high space-time symmetry. Though all proposed
methods have ability to reproduce relativistic equation of motion, the works of D. Felea\cite{Felea01} and
P. Gosselin\cite{Gosselin98} are not Lorentz covariant explicitly. Without introducing
proper time, they treated time and spatial coordinates non-equivalently,
which is contrary to the spirit of relativity. T. Padmanabhan\cite{Padmanabhan94}
employed a quite different approach and introduced proper time, unfortunately it
is still non-covariant. In Pazma's method even the action is non-invariant.\cite{Pazma79}
The only covariant version is originated from Feynman, but it suffers
a difficulty that the physical meanings of the procedure and the conjugated variable
of proper time are unclear. In a sense of that, these path integral schemes
are actually just convenient mathematical techniques for quantum mechanics, and
cannot be treated as the basis for developing a new theory. By contrast, our proposed
formalism is not only Lorentz covariant, but also bears clear physical implications
by developing the rest energy operator $\hat{M}$, which results in the evolution
of states controlled by $\hat{M}$ besides the conventional Hamiltonian, therefore
has the potential to treat variable rest mass problem. Moreover, all previous proposed
formalisms are just defined on coordinate space, It is well known that this kind of path integral
is not always valid, and including phase-space integrals, as in our scheme, is more appropriate.\cite{masujima00}

\section{Discussion}
\label{sec-dis}

It would be interesting to show how our formalism can produce the partial locality for
quantum process, for example, measurements, which is beyond the scope of application of other proposals.
For simplicity, let's assume the measurements are just perturbations to the state, and the corresponding action can be characterized
by a Dirac delta function as $\alpha S'\delta(x_{\mu}-x'_{\mu})$. Then the wave function
with two measurements performed respectively at space-like points of $x'_{u}$ and $x''_{u}$ is given by
\begin{equation}
  \psi(x_{\mu},\tau)=\int_{\mathbb{C}}\psi(x_{\mu}^{0},\tau_{0})\mathcal{D}[x]\exp[\frac{i}{\hbar}\{S_{0}+\alpha S'\delta(x_{u}-x'_{\mu})+\beta S''\delta(x_{u}-x''_{\mu})\}].
\end{equation}
Here we have suppressed unrelated expressions for brief, for example the path integral
in phase space and the detailed information about actions. Parameters $\alpha$ and $\beta$ must be small quantities
to ensure measurements are perturbation. Making Taylor expansion of the exponential to first
order of $\alpha$ and $\beta$, we get
\begin{equation}
\psi(x_{\mu},\tau)\simeq\psi_{0}(x_{\mu},\tau)+\psi'_{D_{1}}(\alpha)+\psi''_{D_{2}}(\beta).
\end{equation}
The first term at the right hand side is the wave function without perturbations and
the last two terms corresponding to influences of measurements whose subscripts $D_{1}$ and $D_{2}$
indicate the domains of definition, which given by
\begin{equation}
\psi'_{D_{1}}(\alpha)=\frac{i}{\hbar}\int_{\mathbb{C}}\mathcal{D}[x]\psi(x_{\mu}^{0},\tau_{0})\alpha S'\delta(x_{u}-x'_{\mu})\exp[\frac{i}{\hbar}S_{0}].
\label{eq:local-1}
\end{equation}
Similar expression holds for $\psi''_{D_{2}}(\beta)$. It is interesting to notice
that the paths with non-zero contributions to Eq.(\ref{eq:local-1}) must pass through
point $x'_{\mu}$. On the other hand, the integral domain $\mathbb{C}$ is determined by
non-spatial condition and $|\mathrm{d}\tau/\mathrm{d}t|\leq 1$ condition, the latter allows us to divide
the integral as
\begin{equation}
\int_{\mathbb{C}}\mathcal{D}[x]=\int_{\mathbb{C}'}\mathcal{D}[x]+\int_{\mathbb{C}''}\mathcal{D}[x]
\end{equation}
where $\mathbb{C}'$ only contains paths with $\mathrm{d}\tau/\mathrm{d}t\leq 1$ and
other paths belong to $\mathbb{C}''$. Evidently, the paths of $\mathbb{C}''$ must
contain contributions arised from $\mathrm{d}\tau/\mathrm{d}t\geq -1$, which describes a
particle moving along reverse time, that is to say, it corresponds to an anti-particle.

At low energy approximation
where no anti-particles are excited, we can ignore the paths belonging to $\mathbb{C}''$ safely.
In this case, at time $(t,\tau)$, the value of wave function are determined by
integral over those paths belonging to $\mathbb{C}'$ and passing trough $x'_{\mu}$
which constrained by non-spatial condition $|\vec x|\leq c(t-t')$, see Fig.\ref{fig:cpi}.
This means that $\psi'_{D_{1}}(\alpha)$ does not be defined around the whole space-time. To enlarge its definition
domain to the whole space-time, we have to use a step function $H(x)$ that has a value of 1 if $x\geq 0$ otherwise 0 to characterize the definition domain $D_{1}$ explicitly.
The result reads
\begin{equation}
\psi'_{\alpha}(x_{\mu},\tau)=H(\Sigma_{\mu}\Delta x_{\mu}\Delta x^{\mu})H(1-\mathrm{d}\tau/\mathrm{d}t)\psi'_{D_{1}}(\alpha),
\end{equation}
where $\Delta x_{\mu}=x_{\mu}-x'_{\mu}$. Similar expression holds for $\psi''_{D_{2}}(\beta)$. Thus if define
$t_{c}=|\vec x'-\vec x''|/2c+(t'+t'')/2$, then when $t\leq t_{c}$ one gets
\begin{align}
  \int\psi'^{*}_{\alpha}(\vec x,t)\psi''_{\beta}(\vec x,t)\mathrm{d}\vec x
  &=\int_{D_{1}\cap D_{2}}\psi'^{*}_{D_{1}}(\alpha)\psi''_{D_{2}}(\beta)\mathrm{d}\vec x\nonumber\\
  &=\int_{\varnothing}\psi'^{*}_{D_{1}}(\alpha)\psi''_{D_{2}}(\beta)\mathrm{d}\vec x\equiv 0.
\end{align}
That is, the influence of two measurements occured at $x'_{\mu}$ and $x''_{\mu}$ are physically uncorrelated.
However, when $t>t_{c}$, it will become correlated since the intersection of $D_{1}$
and $D_{2}$ usually non-empty in this case.

It is quite clear that in this case the propagation of quantum correlation of measurements also has the same speed as light
(if the space-time coordinates of one measurement lies within the influence region
of another measurement, then their results will be correlated, otherwise are independent).
However, if paths in $\mathbb{C}''$ are involved, i.e., with anti-particle excitations,
the segments of a path with $\mathrm{d}\tau/\mathrm{d}t\geq-1$ allow the path
sweeping a larger time interval than actually happened from the point of an observer.
This implies that the permitted spatial region in Fig.\ref{fig:cpi}(b) are broadened,
which would lead to a superluminal propagation of influence region of measurements.
At high energy limits where excited anti-particle can exist permanently,
quantum correlation then will approach globally and instantaneously.

This novel mechanism for propagation of quantum correlation makes it possible to interpret the
experiments on EPR effect\cite{epr35} with a new point of view. A. Aspect's experiments
shows a contradiction between the reality and locality in a physical theory.\cite{aspect81,aspect82} However
since there the locality is defined by the principle of maximum speed of light
(i.e., the velocity of quantum correlation propagation is limited by $c$),
the propagation mechanism of perturbations arised from our formalism releases this
definition and solve the contradiction automatically for massive matters.
As long as energy does not approach infinite, the life-time of excited anti-particles
must be finite, then we have a finite correlation speed and the required isolation
condition for quantum mechanics. If we define this property as locality, which
has the same spirit as the original one---to get rid of action at a distance from
physical theory, then Bell's inequality\cite{bell65} loses its availability to
characterize the relationship of physical reality and locality because we don't
know exact information about anti-particle excitations, as well as the exact
value of correction speed, despite it should larger than the velocity of light usually.
we see that in this case we still have a possibility to preserve the reality and locality
from EPR's experiments. Of course this also leaks a light of hope for hidden
variable theories if they prefer to include anti-particle effects in their framework
to reproduce superluminal mechanism,
though I do not believe it could be a right choice for the truth. It is necessary
to point out that in Aspect's experiments, since correlation between photons are examined,
which corresponding to a singular case in our formalism because photon is massless
and its anti-particle is just itself, the predicted correlation velocity will approach infinite.
This should be the only one case where action at a distance may be possible in our formalism and
results in a maximum violation of Bell's inequality.
However, to eliminate this possibility is far beyond the scope of this paper.
Regards this, we suggest
to perform analogous experiments on massive particles to examine the effects
of anti-particle excitations.

\section{Conclusion}
\label{sec-Conclusion}

Using developed variational principle defined on world lines, we obtained the formal
action for a relativistic system. It permits us to generalize the Feynman's path integral formalism of quantum
mechanics to special relativity case elegantly, which reproduces the Klein-Gordon
and Dirac equation for infinitesimal intervals very well. Some novel understandings
are obtained from this new point of view. Though quantum state can be defined on the whole space-time,
quantum processes are always localized at low energy limits, while large-scale correlations are due
to anti-particles excitations, which makes superluminal corrections become possible
even within the framework of special relativity.
With this mechanism for correlation propagation,
quantum mechanics becomes more completeness and self-consistency. Also
it shows at the first time that we still have opportunity to save the concept
of physical reality without lost the locality of a theory.
Sometimes this kind of correlation mechanism will bring a little inconvenience to calculation. For example, when
considering a process described by a superposition state, one then need to check that if its
sub-states are really correlated or not within studied time interval in order to give physical meaning to the final superimposed state.
At last, it is necessary to note that most discussions in this paper are for single
particle, but the same conclusions hold for many-particles case, which is
rather straightforward to derive and don't need to repeat here.

\section*{Acknowledgments}
The author thankfully acknowledges the help and kind
hospitality from Prof. Ying Chen at the University of Tokyo.

\appendix
\section{Klein-Gordon equation}
Here we would like to derive the Klein-Gordon equation of free particle
explicitly by starting from Eq.(\ref{eq:c-fpi-small-t}), which can be rewritten as
\begin{align}
\psi(x_{\mu},\tau+\varepsilon)=&\int_{\mathbb{C}} \mathrm{d}^{4}x^{0}\int\frac{\mathrm{d}^{4}p}{(2\pi\hbar)^{4}}\,\psi(x^{0}_{\mu},\tau)\nonumber\\
&\times\exp\left[\frac{i}{\hbar}\left\{\sum_{\mu,\nu}p^{\nu}
(x_{\mu}-x^{0}_{\mu})\delta_{\mu\nu}-\varepsilon\cdot \sum_{\nu}\frac{p^{\nu}p_{\nu}}{2m_{0}}\right\}\right],
\label{eq:A-fpi}
\end{align}
here we have used a condition of $\sum_{\mu}p_{\mu}p^{\mu}=m_{0}^{2}c^{2}$ to remove the square root in action, which means that
we are considering a particle with fixed rest mass. Note the 4-momentum in Eq.(\ref{eq:A-fpi}) should be
treated independently when making path integral, namely, it does not satisfy
the mass-energy relation and corresponding to none real physical state. Integral
out the momentum, one gets
\begin{equation}
\psi(x_{\mu},\tau+\varepsilon)=\frac{im_{0}^{2}}{4\pi^{2}\hbar^{2}\varepsilon^{2}}
\int_{\mathbb{C}}\mathrm{d}^{4}(\delta x)\,\psi(x_{\mu}-\delta x_{\mu},\tau)\exp\left[\frac{i}{\hbar}\frac{m_{0}}{2\varepsilon}\sum_{\mu}\delta x_{\mu}\delta x^{\mu}\right].
\label{eq:A-coord-fpi}
\end{equation}
Here use has being made of $\delta x_{\mu}=x_{\mu}-x_{\mu}^{0}$. As $\delta x_{\mu}$
is small and making the Taylor expansion of $\psi$ to its second order, we have
\begin{equation}
  \psi(x_{\mu}^{0})\simeq \psi(x_{\mu})-\sum_{\mu}\delta x_{\mu}\cdot \frac{\partial\psi}{\partial x_{\mu}}+\frac{1}{2\,!}\sum_{\mu,\nu}\delta x_{\mu}\delta x_{\nu}\cdot \frac{\partial^{2}\psi}{\partial x_{\mu}\partial x_{\nu}}+\cdots
\end{equation}
While considering that the integral domain $\mathbb{C}$ is central symmetry
with respect to the origin of coordinates, non-zero contributions to the
integral are just come from terms of $\psi(x_{\mu})+\frac{1}{2}\sum_{\mu}(\delta x_{\mu})^{2}\cdot \frac{\partial^{2}\psi}{\partial x_{\mu}^{2}}$.
Thus we can write the right hand side of Eq.(\ref{eq:A-coord-fpi}) as a sum
of two terms, one depends on $\psi$ and the other depends on
second order derivatives of $\psi$.

Let's calculate the first term firstly, i.e.,
\begin{align}
F.t.&=\frac{im_{0}^{2}}{4\pi^{2}\hbar^{2}\varepsilon^{2}}
\int_{\mathbb{C}}\mathrm{d}^{4}(\delta x)\,\psi(x_{\mu},\tau)\exp\left[\frac{i}{\hbar}\frac{m_{0}}{2\varepsilon}\sum_{\mu}\delta x_{\mu}\delta x^{\mu}\right]\nonumber\\
&=\frac{im_{0}^{2}}{4\pi^{2}\hbar^{2}\varepsilon^{2}}\psi(x_{\mu},\tau)
\int_{\mathbb{C}_{1}}\mathrm{d}(c\delta t)\int_{\mathbb{C}_{2}}\mathrm{d}^{3}(\delta \vec x)\,\exp\left[\frac{i}{\hbar}\frac{m_{0}}{2\varepsilon}\sum_{\mu}\delta x_{\mu}\delta x^{\mu}\right]\nonumber\\
\label{eq:A-ft}
\end{align}
where $\mathbb{C}_{1}$ is given by $[-\infty,-c\varepsilon]\cup[c\varepsilon,\infty]$ and $\mathbb{C}_{2}$
is determined by $|\delta \vec x|\leq c\delta t$ (see Fig.\ref{fig:cpi}).
Define $\alpha=\frac{m_{0}}{2\varepsilon\hbar}$ and integral out the spatial
components, we have
\begin{equation}
  F.t.=\frac{im_{0}^{2}}{4\pi^{2}\hbar^{2}\varepsilon^{2}}\psi(x_{\mu},\tau)
\int_{\mathbb{C}_{1}}\frac{i\pi}{\alpha}e^{-i\alpha (c\delta t)^{2}}\sqrt{\frac{\pi}{i\alpha}\left(e^{i\alpha (c\delta t)^{2}}-1\right)}\,\mathrm{d}(c\delta t).
\end{equation}
Taking account of $e^{i\alpha (c\delta t)^{2}}$ approaches zero when $\varepsilon\rightarrow 0$,
we can make an approximation of the square root simply as $\sqrt{i\pi/\alpha}$ to
get accurate enough result on $O(\varepsilon)$. Thus integral over the time
component and making use of $e^{\pm i\alpha \infty^{2}}\rightarrow 0$, we
finally get
\begin{align}
  F.t.&=\frac{-m_{0}^{2}}{4\pi^{2}\hbar^{2}\varepsilon^{2}}\psi(x_{\mu},\tau)
  (\frac{i\pi}{\alpha})^{2}e^{-i\alpha(c\varepsilon)^{2}/2}\nonumber\\
  &=\psi(x_{\mu},\tau)e^{-i\alpha(c\varepsilon)^{2}/2}\simeq \psi(x_{\mu},\tau)-\frac{im_{0}c^{2}}{4\hbar}\varepsilon\cdot\psi(x_{\mu},\tau).
\end{align}
In the last equality we just keep to the first order of $\varepsilon$.

The calculation of the second term are quite similar,
\begin{align}
S.t.&=\frac{im_{0}^{2}}{8\pi^{2}\hbar^{2}\varepsilon^{2}}
\int_{\mathbb{C}}\mathrm{d}^{4}(\delta x)\,\left[\sum_{\mu}(\delta x_{\mu})^{2}\cdot \frac{\partial^{2}\psi}{\partial x_{\mu}^{2}}\right]\exp\left[i\alpha\sum_{\mu}\delta x_{\mu}\delta x^{\mu}\right]\nonumber\\
&=\frac{im_{0}^{2}}{8\pi^{2}\hbar^{2}\varepsilon^{2}}\left[\sum_{\mu}\partial^{\mu}\partial_{\mu}\psi\right]\left(\frac{i\pi}{\alpha}\right)^{\frac{3}{2}}
\int_{\mathbb{C}_{1}}(c\delta t)^{2}e^{-i\alpha (c\delta t)^{2}}\mathrm{d}(c\delta t).
\label{eq:A-ft2}
\end{align}
Integral term gives
\begin{equation}
  \frac{c\varepsilon}{i\alpha}e^{-i\alpha(c\varepsilon)^{2}}+\frac{1}{i2\alpha}
  \sqrt{\frac{\pi}{i\alpha}}e^{-i\alpha(c\varepsilon)^{2}/2}.
\end{equation}
Regardless the exponential contributions, the first term is in proportion to $O(\varepsilon^{2})$
and the second term in proportion to $O(\varepsilon^{3/2})$. At the first
approximation, we can neglect the first term. Then we have
\begin{align}
S.t.&\simeq\frac{im_{0}^{2}}{8\pi^{2}\hbar^{2}\varepsilon^{2}}\left[\sum_{\mu}\partial^{\mu}\partial_{\mu}\psi\right]\left(\frac{i\pi}{\alpha}\right)^{\frac{3}{2}}
\frac{1}{i2\alpha}
  \sqrt{\frac{\pi}{i\alpha}}e^{-i\alpha(c\varepsilon)^{2}/2}\nonumber\\
  &=\frac{i}{4\alpha}\,e^{-i\alpha(c\varepsilon)^{2}/2}\sum_{\mu}\partial^{\mu}\partial_{\mu}\psi
  \simeq\frac{i\hbar\varepsilon}{2m_{0}}\sum_{\mu}\partial^{\mu}\partial_{\mu}\psi.
\end{align}

In this way we get
\begin{equation}
\psi(x_{\mu},\tau+\varepsilon)=\psi(x_{\mu},\tau)-\frac{im_{0}c^{2}}{4\hbar}\varepsilon\cdot\psi(x_{\mu},\tau)
+\frac{i\hbar\varepsilon}{2m_{0}}\sum_{\mu}\partial^{\mu}\partial_{\mu}\psi(x_{\mu},\tau).
\end{equation}
Expand the left hand term to first order of $\varepsilon$ and comparing with
right hand terms, one has one trivial identical equality and
\begin{equation}
\frac{\partial\psi}{\partial\tau}+\frac{im_{0}c^{2}}{4\hbar}\psi=
\frac{i\hbar}{2m_{0}}\sum_{\mu}\partial^{\mu}\partial_{\mu}\psi.
\end{equation}
for first order of $\varepsilon$. Define $\phi=e^{\frac{im_{0}c^{2}}{4\hbar}\tau}\psi$, one can rewrite above
equation as
\begin{equation}
  \frac{\partial\phi}{\partial\tau}=\frac{i\hbar}{2m_{0}}\sum_{\mu}\partial^{\mu}\partial_{\mu}\phi,
\end{equation}
or,
\begin{equation}
  -\hbar^{2}\frac{\partial^{2}\phi}{\partial\tau^{2}}=-\frac{\hbar^{2}c^{2}}{4}\sum_{\mu}\partial^{\mu}\partial_{\mu}\phi,
\end{equation}
i.e., $-\hbar^{2}\partial^{2}\phi/\partial\tau^{2}=\hat{M}^{2}\phi$ with $\phi=\phi(x_{\mu})e^{-\frac{im_{0}c^{2}}{2\hbar}\tau}$.
For space-time component $\phi(x_{\mu})$, it satisfies
\begin{equation}
  m_{0}^{2}c^{2}\phi(x_{\mu})=-\hbar^{2}\sum_{\mu}\partial^{\mu}\partial_{\mu}\phi(x_{\mu}),
\end{equation}
which is exactly the Klein-Gordon equation.

\section{Non-relativistic approximation}

To obtain the non-relativistic approximation, it is natural to set light speed $c\rightarrow\infty$.
Then we have $\mathrm{d}\tau/\mathrm{d}t\rightarrow 1$ at classical limit for massive
particles when time-arrow (i.e., the direction of proper time is identical with
time) is employed. This indicates that the domain $\mathbb{C}_{1}$ in figure \ref{fig:cpi}(a)
becomes a straight line which is given by $\tau=t_{0}+t$, and the proper time is
dependent and superfluous. The summation lattice in figure \ref{fig:cpi}(b)
also becomes equivalent to the standard path integral of non-relativistic quantum mechanics
because $|\vec x|\leq c\cdot\delta t\rightarrow\infty$.

Under this approximation, the action can be expressed as
\begin{equation}
  \int_{\tau_{i}}^{\tau_{f}}L(x_{\mu},\dot{x}_{\mu})\,\mathrm{d}\tau\rightarrow\int_{t_{i}}^{t_{f}}L(\vec x,\frac{{\mathrm{d}\vec x}}{\mathrm{d}t})\,\mathrm{d}t+const.,
\end{equation}
namely, the non-relativistic action between $t_{i}$ and $t_{f}$ plus a constant because
$\int\mathrm{d}\tau=\int\mathrm{d}t$ and $\mathrm{d}x_{\mu}/\mathrm{d}\tau=(\mathrm{d}t/\mathrm{d}\tau)\cdot\mathrm{d}x_{\mu}/\mathrm{d}t=\mathrm{d}x_{\mu}/\mathrm{d}t$.
Take a free particle as example, the Lagrangian is
\begin{equation}
  L=\frac{m_{0}}{2}\sum_{\mu}\dot{x}_{\mu}\dot{x}_{\mu}\rightarrow\frac{m_{0}c^{2}}{2}
  -\frac{m_{0}\vec v^{2}}{2}=\frac{m_{0}c^{2}}{2}+L_{nr}
\end{equation}
where subscript $nr$ refers to standard non-relativistic Lagrangian.
Then the action can be written as $S\rightarrow(t_{f}-t_{i})m_{0}c^{2}/2+S_{nr}$.
Therefore from Eq.(\ref{eq:c-fpi}) we have the transition probability amplitude
between two quantum states as
\begin{align}
\langle \vec x, t_{f}&|\vec x,t_{i}\rangle=\lim_{n\rightarrow\infty \atop n\cdot\delta t=t_{f}-t_{i}}
\int_{t_{i}}^{t_{f}}\prod_{k=1}^{n-1}\frac{\mathrm{d}t_{k}}{(2\pi\hbar)^{1/2}}
\prod_{k=0}^{n-1}\frac{\mathrm{d}E_{k}}{(2\pi\hbar)^{1/2}}\nonumber\\
&\times\int_{x(t)=x(t_{i})}^{x(t)=x(t_{f})}\prod_{k=1}^{n-1}\frac{\mathrm{d}^{3}x_{t_{k}}}{\left(2\pi\hbar\right)^{3/2}}
\int\prod_{k=0}^{n-1}\frac{\mathrm{d}^{3}p_{t_{k}}}{\left(2\pi\hbar\right)^{3/2}}
\cdot e^{\frac{i}{\hbar}S_{nr}}\cdot e^{\frac{i}{\hbar}\frac{m_{0}c^{2}}{2}(t_{f}-t_{i})}.
\end{align}
Taking out the constant $v^{n}e^{\frac{i}{\hbar}\frac{m_{0}c^{2}}{2}(t_{f}-t_{i})}$
where
\begin{equation}
  v^{n}=\lim_{n\rightarrow\infty \atop n\cdot\delta t=t_{f}-t_{i}}(\delta t)^{n-1}\prod_{k=0}^{n-1}\int\frac{\mathrm{d}E_{k}}{2\pi\hbar},
\end{equation}
we get the transition function as
\begin{equation}
\langle \vec x, t_{f}|\vec x,t_{i}\rangle\propto\lim_{n\rightarrow\infty \atop n\cdot\delta t=t_{f}-t_{i}}
\int_{x(t)=x(t_{i})}^{x(t)=x(t_{f})}\prod_{k=1}^{n-1}\frac{\mathrm{d}^{3}x_{t_{k}}}{\left(2\pi\hbar\right)^{3/2}}
\int\prod_{k=0}^{n-1}\frac{\mathrm{d}^{3}p_{t_{k}}}{\left(2\pi\hbar\right)^{3/2}}
\cdot e^{\frac{i}{\hbar}S_{nr}}.
\end{equation}
The right side is exactly the standard non-relativistic path integral formalism for free particles.\cite{masujima00}
When potentials due to pure gauge fields are present, the conclusion still holds
where instead Eq.(\ref{eq:action-2}) should be used.

\end{document}